%Paper: hep-th/9309091
%From: Peter Landshoff <P.V.Landshoff@damtp.cambridge.ac.uk>
%Date: Thu, 16 Sep 93 14:34:28 BST

%Plain tex file, including one figure. In addition to pictex, the file
%calls banffpic.tex and banff.a

 \headline={\ifnum\pageno=1\firstheadline\else
 \ifodd\pageno\rightheadline \else\leftheadline\fi\fi}
 \def\firstheadline{\hfil}
 \def\rightheadline{\hfil}
 \def\leftheadline{\hfil}
 	\footline={\ifnum\pageno=1\firstfootline\else\otherfootline\fi}
 \def\firstfootline{\rm\hss\folio\hss}
 \def\otherfootline{\hfil}
  %%REFERENCES
%     \defref\label{text}
% generates a number, assigns it to \label, generates an entry.
% To list the refs,  \listrefs
% (Extracted and adapted from harvmac.tex by P Ginsparg)

\font\rfont=cmr10 at 10 true pt
\def\ref#1{$^{\hbox{\rfont {#1}}}$}
\catcode`\@=11 % This allows us to modify PLAIN macros.

\def\nolabels{\def\wrlabeL##1{}\def\eqlabeL##1{}\def\reflabeL##1{}}
\def\writelabels{\def\wrlabeL##1{\leavevmode\vadjust{\rlap{\smash%
{\line{{\escapechar=` \hfill\rlap{\sevenrm\hskip.03in\string##1}}}}}}}%
\def\eqlabeL##1{{\escapechar-1\rlap{\sevenrm\hskip.05in\string##1}}}%
\def\reflabeL##1{\noexpand\llap{\noexpand\sevenrm\string\string\string##1}}}
\nolabels
\global\newcount\refno \global\refno=1
\newwrite\rfile
\def\defref{$^{{\hbox{\rfont \the\refno}}}$\nref}
\def\nref#1{\xdef#1{\the\refno}\writedef{#1\leftbracket#1}%
\ifnum\refno=1\immediate\openout\rfile=refs.tmp\fi
\global\advance\refno by1\chardef\wfile=\rfile\immediate
\write\rfile{\noexpand\item{#1\ }\reflabeL{#1\hskip.31in}\pctsign}\findarg}
%	horrible hack to sidestep tex \write limitation
\def\findarg#1#{\begingroup\obeylines\newlinechar=`\^^M\pass@rg}
{\obeylines\gdef\pass@rg#1{\writ@line\relax #1^^M\hbox{}^^M}%
\gdef\writ@line#1^^M{\expandafter\toks0\expandafter{\striprel@x #1}%
\edef\next{\the\toks0}\ifx\next\em@rk\let\next=\endgroup\else\ifx\next\empty%
\else\immediate\write\wfile{\the\toks0}\fi\let\next=\writ@line\fi\next\relax}}
\def\striprel@x#1{} \def\em@rk{\hbox{}}
\def\lref{\begingroup\obeylines\lr@f}
\def\lr@f#1#2{\gdef#1{\defref#1{#2}}\endgroup\unskip}
\newwrite\lfile
{\escapechar-1\xdef\pctsign{\string\%}\xdef\leftbracket{\string\{}
\xdef\rightbracket{\string\}}}

\def\writestop{\def\writestoppt{\immediate\write\lfile{\string\p
ageno%
\the\pageno\string\startrefs\leftbracket\the\refno\rightbracket%
\string\def\string\secsym\leftbracket\secsym\rightbracket%
\string\secno\the\secno\string\meqno\the\meqno}\immediate\closeout\lfile}}
\def\writestoppt{}\def\writedef#1{}
\catcode`\@=12 % at signs are no longer letters

 \font\twelvebf=cmbx10 scaled\magstep 1
 \font\twelverm=cmr10 scaled\magstep 1
 \font\twelveit=cmti10 scaled\magstep 1

 \font\tenbf=cmbx10
 \font\tenrm=cmr10
 \font\tenit=cmti10

 \parindent=1.5pc
 \hsize=6.0truein
 \vsize=8.5truein
 \nopagenumbers

% \centerline{\tenbf }
 \baselineskip=22pt
 \centerline{\bf CANONICAL QUANTISATION OF THERMAL}
 \baselineskip=16pt
 \centerline{\tenbf GAUGE THEORIES}
 %\centerline{\ninerm (For 20\% Reduction to 8.5 $\times$ 6 in Trim Size)}
 \vglue 0.8cm
 \centerline{\tenrm P V LANDSHOFF}
 \baselineskip=13pt
 \centerline{\tenit DAMTP, Cambridge University}
 \baselineskip=12pt
 \centerline{\tenit Cambridge CB3 9EW, United Kingdom}
 \vglue 0.3cm
% \centerline{\tenrm and}
% \vglue 0.3cm
% \centerline{\tenrm SECOND AUTHOR'S NAME}
%\baselineskip=13pt
% \centerline{\tenit Group, Company, Address, City, State ZIP/Zone, Country}
% \vglue 0.8cm
 \centerline{\tenrm ABSTRACT}
 \vglue 0.3cm
 {\rightskip=3pc
  \leftskip=3pc
  \tenrm\baselineskip=12pt\noindent
Canonical quantisation gives a new and convenient finite-temperature
perturbation theory in covariant gauges, and solves the problem of the
zero-frequency mode in the temporal gauge.
 \vglue 0.6cm}
\def\tr{\rm tr}
 \vfil
 \twelverm
 \baselineskip=14pt
 \leftline{\twelvebf 1. Introduction}
 \vglue 0.4cm
 %\vglue 0.3cm
Canonical quantisation is the most basic method of deriving the Feynman
rules for thermal gauge theories.  In this talk, I review work that I
have done with A. ~Rebhan\defref\rebhan{
P V Landshoff and A Rebhan, Nuclear Physics B383 (1992) 607 and preprint
DAMTP 93-06, TUW-93-03 (Nuclear Physics B, in press)
} and with K.~A. James\defref\james{
K A James and P V Landshoff, Physics Letters B251 (1990) 167
} on the following:
\vglue 0.3cm
\noindent$\bullet$ \underbar{Covariant gauges in the real-time
formalism\ref{\rebhan}}\hfill\break

We have derived Feynman rules that can be simpler to use than the
conventional ones, and which avoid the problems that have been
encountered in demonstrating the gauge%
-independence of certain
physical quantities.
\vglue 0.3cm
\noindent$\bullet$ \underbar{The temporal-axial gauge in the imaginary-time
formalism\ref{\james}}\hfill\break

At zero temperature, the propagator has a term with a double pole at
$k = 0$.  Naively, it has been expected\defref\kajantie{
K Kajantie and J Kapusta, Annals of Physics 160 (1985) 477
} that at finite
temperature the integral of this should be replaced with
$\sum_n\,{1/(n^2 \pi^2 T^2)}$.  We have shown how to avoid the apparent
difficulty with the $n=0$ term in this sum.
\vglue 0.3cm
\noindent I am going to try to make my description as simple as possible; full
details may be found in our original papers.
\vglue .6cm
\leftline{\twelvebf 2. Covariant gauges in the real-time formalism}
\vglue 0.3cm
 Although the heading of this section explicitly refers to covariant
gauges, most of what I have to say applies to {\twelveit any} gauge.

The reason that the quantisation of gauge theories involves
subtleties, particularly at finite temperature, is that the Hilbert
space includes unphysical states. The thermal average of an
observable, derived by considering a grand ensemble, requires one to
take expectation values only in the physical states.  So if one wants
to write this thermal average as a trace, one needs to include a
projection operator ${\cal P}$ that removes the unphysical states:

$$<Q> = Z^{-1}\,\tr\,\left (e^{-\beta H}\,Q\,{\cal P}\right) \eqno(1)$$

\noindent where the operators are in the Heisenberg picture.  Some of the
important steps in the derivation of the Feynman rules for scalar
thermal field theory rely on the identity

$$ \tr\,(AB) = \tr\,(BA). \eqno (2) $$

\noindent For the case of a gauge theory, the presence of the projection
operator changes things: it is not true that

$$ \tr\,(AB{\cal P}) = \tr\, (BA{\cal P}). \eqno (3) $$

\noindent One sees the first effect of this when one goes over to the
interaction picture:

$$
\langle Q\rangle\ =\ Z^{-1}\,\tr\,\left(e^{-\beta H_0}\ U\,(t_0 -
i\beta,\,t)\,Q\,(t)\,U\,(t,
t_0)\,{\cal P}\right). \eqno{(4)}
$$

\noindent This follows directly from the usual definition of the
interaction-picture state evolution operator $U$. The time $t_0$
must\ref{\james}
be the time at which the interaction picture coincides with the
Heisenberg picture.  In the scalar-field case, where there is no
projection operator ${\cal P}$, one can show from the trace identity
Eq.\ (2) that $t_0$ can be any time.

For the real-time formalism, one lets $t_0 \rightarrow - \infty$.  One
assumes that one may switch off the gauge coupling very slowly: $g
\rightarrow g\,e^{-\epsilon|t|}$.  This has no effect at $t=0$, and so
will not change the thermal average (1) at $t=0$, which depends only on
the configuration of the system now, independently of its past
history.  What it does, is to change the differential equation that
$U$ satisfies, but only by terms of order $\epsilon$. It is
well-established that it is safe to neglect these when $\epsilon
\rightarrow 0$.  The consequence of the switching off of
 $g$ is that
the interaction-picture physical states are simple: they may be
assumed to contain only transverse gluons.  So we may regard the two
transverse components of the gluon field as physical, while the other
two components, and the Faddeev-Popov ghost, are unphysical.

Now in the real-time formalism, the propagator for any field is a $2
\times 2$ matrix\defref\niemi{
L V Keldysh, Sov Phys JETP 20 (1965) 1018\hfill\break
A Niemi and G W Semenoff, Nuclear Physics B230 (1984) 181
} consisting of the zero-temperature vacuum part, plus a
thermal part. The thermal part is proportional to $\delta\,(k^2)$.
This is because it represents the effect of the real on-shell gluons
in the heat bath.

The heat bath contains only the physical gluons.  Thus it is immediate
that only their propagator need have a thermal part.  The unphysical
fields can remain frozen, with no thermal parts in their propagators.
This contrasts with the conventional formalism, where the unphysical
gluon propagators are given thermal parts too, and their effects are
then cancelled by the ghost propagator also having a thermal part.

The thermal propagator for the gluon field may be calculated in
covariant gauges by Gupta-Bleuler quantisation\defref\gupta{
T Kugo and I Ojima, Physics Letters 73B (1978) 459\hfill\break
I Lazzizzera, Proc Workshop on Nonstandard Gauges, Vienna 1985
(Springer)\hfill\break
P V Landshoff and P van Nieuwenhuizen, preprint DAMTP 93-33, ITP-SB-93-38
}. With the Keldysh time
path\ref{\niemi}, it is defined as

$$ iD^{\mu\nu}\,(x)\ =\ \left[\matrix{\left\langle T\,A^\mu\,(x)\
A^\nu\,(0)\right\rangle& \left\langle A^\mu\,(0)\
A^\nu\,(x)\right\rangle\cr
\left\langle A^\mu\,(x)\ A^\nu\,(0)\right\rangle&\left\langle\bar{T}A^\mu\,(x)\
A^\nu\,(0)\right\rangle}\right ].\eqno{(5)}$$

\noindent This is straightforward to calculate. I choose to write the result as

$$ D\ =\ M \hat {D} M \eqno {(6a)}$$

\noindent where $\hat{D}$ is the diagonal matrix

$$ \hat{D}\ =\ \left[\matrix{{1}\over{k^2+i\epsilon}&0\cr
0&{-1}\over{k^2-i\epsilon}}\right].\eqno {(6b)}$$

\noindent For the transverse physical fields, the matrix $M$ is

$$M_T\ =\ \sqrt{n\left(|k_0|\right)}\ \left[\matrix{
e^{{1\over{2}}\beta|k_0|}&e^{-{{1}\over{2}}\beta k_0}\cr
e^{{{1}\over{2}}\beta k_0}&e^{{1\over{2}}\beta|k_0|}}\right]\eqno{(7a)}$$

\noindent with $n$ the Bose distribution. For the unphysical fields,
$M_T$ must be replaced by its $T\rightarrow 0$ limit:

$$M_0\ =\ \left[\matrix{1&\theta(-k_0)\cr
\theta(k_0)&1}\right].\eqno{(7b)}$$

\noindent The physical field has the same tensor structure in all
gauges, namely $T^{\mu\nu}$ with

$$\eqalign{T^{\mu 0}&=\ 0\ =\ T^{0\nu}\cr
T^{ij}&=\ -\delta^{ij}+{{k^ik^j}\over{{\bf k}^2}}.}\eqno{(8)}$$

\noindent The tensor structure of the unphysical gluon field propagator varies
from gauge to gauge. In covariant gauges, it involves the gauge
parameter. Because it does not have a thermal part, the formalism
avoids some of the problems that are usually encountered with
gauge-parameter dependence\defref\baier{
R Baier, G Kunstatter and D Schiff, Physical Review D45 (1992) R4381
}. Although unphysical quantities may still be gauge-dependent, because
of the gauge dependence of the zero-temperature part of the propagator,
we do not need any subtle techniques to deal with\defref\rebhan2{
A Rebhan, Physical Review D46 (1992) 4779
} aggravated infrared behaviour of unphysical modes.

\vglue .6cm
\leftline {\twelvebf 3. Applications}
\vglue 0.4cm
By not heating the unphysical fields, we simplify calculations.  An
example is the hard thermal loop, which is the leading contribution at
high temperature to the one-loop gluon self-energy. Because the ghost
field is not heated, only gluon loops contribute. By manipulating the
integrands it is straightforward to show that the hard thermal loop
is gauge invariant, without explicitly calculating the Feynman
integrals.

The leading temperature-dependent contribution to the imaginary part
of the gluon self-energy similarly does not have a ghost part: it is
calculated from the gluon loop with both gluons heated.  Because the
heated part of the gluon propagator is the same in all gauges, it is
trivial that the answer is gauge invariant. The resulting damping
constant is the same, in all gauges, as was originally calculated with
the Coulomb gauge\defref\heinz{
U Heinz, K Kajantie and T Toimela, Annals of Physics 176 (1987) 218
}.

So if our method of calculation had been invented before the
conventional one, the need would not have been so obvious for Braaten-Pisarski
resummation\defref\braaten{
E Braaten and R D Pisarski, Phys Rev Lett 64  (1990) 1338
}.  One might ask, therefore, why such resummation is needed
in our approach.

A partial answer to this question comes from looking at multiple
self-energy insertions in the gluon propagator.  In the conventional
formalism, the bare propagator $D$ may be written in the form (6a),
and also the self-energy is

$$
\Pi = M^{-1} \hat{\Pi} M^{-1} \eqno (9)
$$

\noindent with $\hat {\Pi}$ diagonal and the same matrix $M$: in both cases $M
=
M_T$. So when two self-energy insertions are made in the propagator,
the result $D\Pi D\Pi D$ is just
$M\hat{D}\hat{\Pi}\hat{D}\hat{\Pi}\hat{D}M$.
Hence the two entries $(k^2
\pm i\epsilon^{-1})^{-1}$ in $\hat{D}$, given in Eq.\ (6b), do not get
multiplied
together.  In our formalism, however, we have a mixture of matrices
$M_T$ and $M_0$, so that this is no longer true, and in applications
there is a risk that when the dressed propagator is integrated the two
poles pinch the contour of the integration.

It may be that in the calculation of a physical quantity the
numerators and $\delta$-functions that also appear under the integral
kill the potential pinch.  We have shown that this is what happens
with the 3-loop pressure.

However, in the unresummed propagator itself, the pinches are
certainly there.  But resummation removes them.  For example, the
hard-thermal-loop contribution to the resummed spatially-longitudinal
propagator is given by

$$
D^{-1}_{L}\ =\ M_0^{-1}\,\hat{D}^{-1}\,M_0^{-1}\ -\
M^{-1}_T\,\hat{\Pi}_L\,M^{-1}_T. \eqno
(10a)
$$

\noindent Simple algebra, using the explicit forms (7) for $M_T$ and $M_0$,
gives

$$
D_L^{-1}\ =\
M^{-1}\,\left[\matrix{k^2-\Pi_L+i\epsilon&2i\epsilon\,e^{-\beta
k_0}\,\theta(k_0)\cr
2i\epsilon\,e^{\beta
k_0}\,\theta(-k_0)&-k^2+\Pi^{\ast}_L+i\epsilon}\right]\,M^{-1}.\eqno{(10b)}
$$

\noindent When $\epsilon \rightarrow 0$, the off-diagonal terms vanish and the
pinch disappears. It is not completely trivial that one may safely
take this limit, though careful argument does justify it. Thus,
although the bare spatially-longitudinal propagator is frozen at
$T=0$, the self-energy insertion heats it to temperature $T$, so that
after resummation it is similar in structure to the transverse
propagator. The mechanism by which this occurs was first discussed by
Weldon\defref\weldon{
H A Weldon, Physical Review D45 (1992) 352
}. But note that, in the hard-thermal-loop approximation, the
remaining component of the gluon propagator, and the ghost, remain
frozen.

Beyond the hard-thermal-loop approximation, it is possible that things
are more complicated.
\vglue .6cm
\leftline{\twelvebf 4. $A^0 = 0$ gauge in the imaginary-time
formalism}
\vglue 0.3cm
With canonical quantisation, the $(1,1)$ component of the longitudinal
gluon propagator in $A^0=0$ gauge is found to be

$$D^{11}\,(\tau,{\bf k})\ =\ i|\tau|\,\theta(k_3\tau)\eqno (11)$$

\noindent where $\tau=it$. The $k_3$ appears because of the particular
quantisation procedure; it may be replaced with any other component of
${\bf k}$. Because Eq.\ (11) is independent of temperature, clearly it
does not satisfy KMS periodicity. The reason for this is that the
projection operator ${\cal P}$ in Eq.\ (1) prevents the use of the
trace identity Eq.\ (2) that is needed for the usual demonstration of
periodicity. The consequence is that both bosonic and fermionic
Matsubara frequencies are needed for the longitudinal propagator:

$$\eqalign{D_L^n\,({\bf k})\ =&{1\over 2}i
\int^{-\beta}_{\beta}d\tau\,D_L^{11}(\tau,{\bf k})\,e^{i\pi
n\beta\tau}\cr
=&\left\{\eqalign{1/4T^2\hskip .5in&\hskip .5in n=0\cr
-i\epsilon(k_3)/2\pi n^2T^2&\hskip .5in n\ {\rm even}\not= 0\cr
\dots\hskip .5in&\hskip .5in n\ {\rm odd}}\right.}\eqno (12)$$

\noindent Hence the temperature has regulated the naively-expected
$1/n^2$ singularity at $n=0$. I have not written the expression for
$D_L^n$ explicitly when $n$ is odd, because in practice it is easier
not to make a frequency summation, but to integrate directly in
$t$-space.

The result Eq.\ (11) is simple enough, but there is a complication.
The imaginary-time formalism requires the time $t_0$ in Eq.\ (4) to be
finite. But this is the time at which the interaction picture
coincides with the Heisenberg picture. So now we cannot switch off the
QCD interaction when we define the interaction-picture physical
states. They are defined in two equivalent ways, either by requiring
that matrix elements of the Gauss-law operator vanishes, or from\ref{\gupta}

$$Q_{BRST}\,\left.\bigm|\ \right\rangle\ =\ 0.\eqno (13)$$

\noindent The result is that the physical states are no longer those
that contain only transverse gluons, but some mapping of these states:

$$\left.\bigm|\ \right\rangle\ =\ \left.\Lambda_{t_0}\,\bigm|{\rm
TRANSVERSE} \right\rangle\eqno (14)$$

\noindent at $t=t_0$, where $\Lambda_{t_0}$ is some functional of the $t=t_0$
field. Consequently, instead of Eq.\ (4) we have

$$\langle Q\rangle\ =\ Z^{-1}\,\tr\,\left(e^{-\beta H_0}\
\Lambda^{\dag}_{t_0-i\beta}\,U\,(t_0 -
i\beta,\,t)\,Q\,(t)\,U\,(t,t_0)\,\Lambda_{t_0}\right) \eqno{(15)}
$$

\noindent where now the trace is taken only over the subspace of the
transverse states. Note that $\Lambda^{\dag}_{t_0-i\beta}$ appears, rather than
$\Lambda^{\dag}_{t_0}$, because it has been moved past the $e^{-\beta H_0}$.

We may expand $\Lambda_{t_0}$ in a power series:

$$\Lambda\ =\ 1\ +\ g\lambda_1\ +\ g^2\lambda_2\ +\dots\eqno (16)$$

\noindent where the series begins with 1 because when $g\rightarrow0$ the
transverse states themselves are physical. The coefficients $\lambda$
may be calculated from Eq.\ (13). The result for $\lambda_1$ is
trilinear in the gluon field, so that $\lambda_1$ is effectively a
3-gluon vertex additional to the usual QCD vertex. We have tested this
by calculating the two-loop pressure. The diagrams are shown in Figure
1, where the spot denotes the $\lambda_1$ vertex. It is interesting
that the last three diagrams just cancel each other; I do not know
why. The result agrees with previous calculations.
\vskip 0.7truecm
%This file is banffpic.tex
\input pictex.tex
\linethickness=1truemm
\setcoordinatesystem units <10truemm, 10truemm> point at 0 0
\beginpicture

\newbox\picA
\setbox\picA=\hbox{\beginpicture
%\savelinesandcurves on "banff.a"
\replot "banff.a"
\setcoordinatesystem point at 0 0
\setsolid
\plot 0 0 3 0 /
\ellipticalarc axes ratio 3:1 360 degrees from 0 0 center at 1.5 0
\endpicture}

\put{\copy\picA} at 0.15 0
\put{\copy\picA} at 4.15 0
\put{\copy\picA} at 8.15 0
\put{\copy\picA} at 12.15 0

\font\bigmath=cmsy10 scaled\magstep5
{\bigmath
\put{$\bullet$} at 4.05 0
\put{$\bullet$} at 11.05 0
\put{$\bullet$} at 12.05 0
\put{$\bullet$} at 15.05 0
}
\endpicture

\vskip 0.7truecm
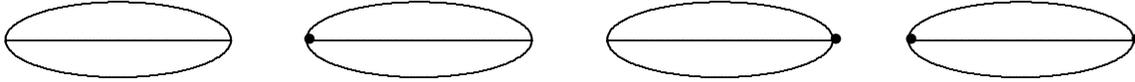
{\rm Figure 1: contributions to the pressure}

\vglue .6cm
\leftline {\twelvebf 5. Summary}
\vglue 0.4cm
\noindent For the real-time formalism, in any covariant gauge,
\vglue 0.3cm
\item{$\bullet$} at bare-propagator level, only the transverse
gluons need to be heated\hfill\break
\vglue 0.1cm
\item{$\bullet$} Braaten-Pisarski resummation heats also the
spatially-longitudinal gluons\hfill\break
\vglue 0.1cm
\item{$\bullet$} the remaining component of the gluon field, and the
Faddeev-Popov ghost, remain frozen at zero temperature\hfill\break
\vglue 0.1cm
\item{$\bullet$} certain gauge-invariance problems of the
conventional formalism  are avoided.\hfill\break
\vglue 0.3cm
\noindent For the imaginary-time formalism,
\vglue 0.3cm
\item{$\bullet$} we know how to calculate in the gauge $A^0=0$.\hfill\break

\noindent However, although I am confident that the procedure I have outlined
for this is correct, it may be that there is a simpler one.

\medskip\immediate\closeout\rfile\writestoppt
\baselineskip=14pt{\bigskip{\twelvebf References}}\bigskip{\frenchspacing%
\parindent=20pt\escapechar=` \input refs.tmp\bigskip}\nonfrenchspacing
\bye